\documentclass[letterpaper, 10pt, conference]{ieeeconf}      
\IEEEoverridecommandlockouts                              
\usepackage[utf8]{inputenc}
\usepackage[T1]{fontenc}
\usepackage{float}
\usepackage{graphicx}
\usepackage{amsmath}
\usepackage{algorithm}
\usepackage[noend]{algpseudocode}
\usepackage{multicol}
\usepackage{afterpage}
\usepackage{amssymb}
\usepackage{amsfonts} 
\usepackage{url}
\usepackage{bbding}
\usepackage{graphicx}
\usepackage{hyperref}
\usepackage{xcolor}
\usepackage{caption}
\usepackage{subcaption}

\title{\LARGE \bf
World War III Analysis using Signed Social Networks
}

\author{Ranjana Roy Chowdhury$^{1}$\ and Shivam Gupta$^{1}$\ and\ Sravanthi Chede$^{1}$

\thanks{*All authors contributed equally and are arranged alphabetically.}
\thanks{$^{1}$Department of Computer Science and Engineering, Indian Institute of Technology (IIT), Ropar, Rupnagar, Punjab, India \small{Correspondence at shivam.20csz0004@iitrpr.ac.in}
     }
}

\begin{document}

\maketitle

\begin{abstract}

In the recent period of time with a lot of social platforms emerging, the relationships among various units can be framed with respect to either positive, negative or no relation. These units can be individuals, countries or others that form the basic structural component of a signed network. These signed networks picture a dynamic characteristic of the graph so formed allowing only few combinations of signs that brings the structural balance theorem in picture. Structural balance theory affirms that signed social networks tend to be organized so as to avoid conflictual situations, corresponding to cycles of unstable relations. The aim of structural balance in networks is to find proper partitions of nodes that guarantee equilibrium in the system allowing only few combination triangles with signed edges to be permitted in graph. Most of the works in this field of networking have either explained the importance of signed graph or have applied the balance theorem and tried to solve problems. Following the recent time trends with each nation emerging to be superior and competing to be the best, the probable doubt of happening of WW-III(World War-III) comes into every individuals mind. Nevertheless, our paper aims at answering some of the interesting questions on World War-III. In this paper we have worked with the creation of a signed graph picturing the World War-III participating countries as nodes and have predicted the best possible coalition of countries that will be formed during war. Also, we have visually depicted the number of communities that will be formed in this war and the participating countries in each communities. Our paper involves extensive analysis on the various parameters influencing the above predictions and also creation of a new data-set of World War -III that  contains the pairwise relationship data of countries with various parameters influencing prediction. This paper also validates and analyses the predicted result.\\

\end{abstract}

\section{INTRODUCTION}

Social networking allows like-minded individuals to be in touch with each other using websites and web-based applications. Facebook, MySpace, Twitter, and LinkedIn are examples of social networking sites. Online social media serves as a platform to show these relationships, whether friendly or unfriendly, like or dislike, agreement or dissension, trust or distrust. These types of interactions lead to the emergence of Signed Social Networks (SSNs)\cite{DBLP:journals/fgcs/LiuQNW20} where positive sign represents friend, like, trust, agreement and negative sign represents foe, dislike, distrust and disagreement. The richness of SSNs consists of a mixture of both positive and
negative interactions, co-existing in a single structure. 

The main motive in SSNs is to examine the interplay between positive and negative links in social media as well as to have a network or network partitions that look at global balance rather than local balance. Balance here refers to fact that  certain configurations of positive and negative edges will be favored and others are disfavored. Such situation give rise to the application of 'Structural Balance Theory' which considers the possible ways in which triangles between individuals can be signed and posits that the network alters over time toward particular structural forms \cite{DBLP:journals/corr/abs-1901-06845}. It means that only a certain combination of signs in the edges of a triangle permutation depicting relationships will be allowed. Fig 1.A and Fig 1.B depicts the balance as well as unbalanced combination of triangles taken in a SSNs. The aim of network structural balance theorem is to find proper partitions of nodes that guarantee equilibrium in the SSNs.\\

\begin{figure}[H]
    \centering
    \includegraphics[width=8cm]{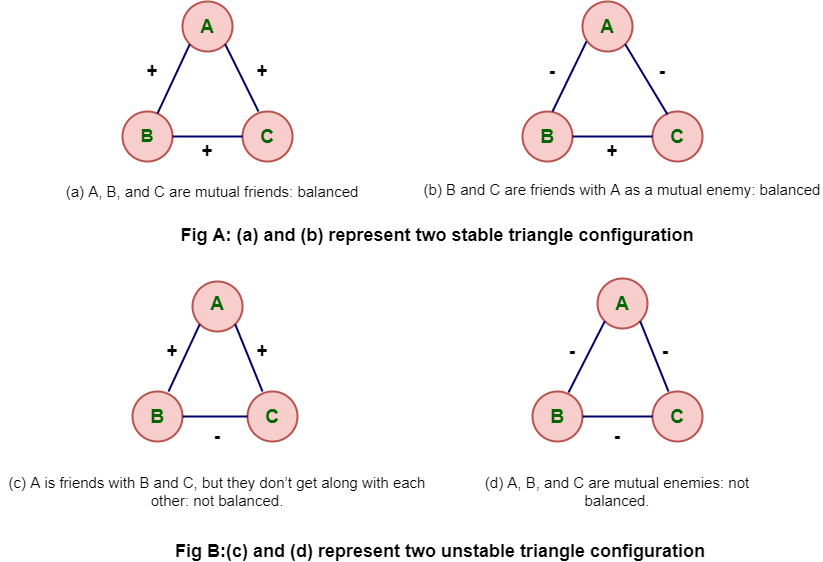}
    \textbf{\caption{Figure representing structural balance in triangles}} 
    \label{fig:my_label}
\end{figure}

Thus it has been always a tremendous area of interest for researchers to predict the outcome of events that are modelled using SSNs. Among many challenging and interesting problems, one of the most inquisitive questions that strikes into each of us is that, will there be another war within few forthcoming years?\\
Considering the present situation where the temperament of each country lies in boosting its power and efficiency, it reigns in the parliamentary constitution of each country to be independent in terms of all factors that makes the country stable as well as emerge as a renowned nation overall. But, due to some unfavourable circumstances some of the powerful countries of this world reigns to overpower others and establish its superiority in a rather non-friendly manner. Also, there are still few countries who are colonies of powerful countries that wish to be independent. 
Today all nations are in a rat-race to acquire more power and leadership. To achieve this every nation is trying to control the resources of other nations. Also, gaining control is not sufficient as each nation needs to ensure its own survival as well. The race for power often leads to fights between two nations (namely superpowers). These fights can turn violent and get interlinked with other nations. Thus finally resulting in a world war between nations in order to gain power. The shift in balance of power and its related study has always attracted researchers from across the globe. A significant effort has been made to study international relations and bonds between nations. The relationships between nations form a complex system. One can use network models to get an abstract representation of such systems. The relationship among nations can either be negative (conflicting) tie or positive (harmonious) tie.\\
These conditions often heat up the beam and will eventually lead to knocking up of the upcoming World War-III. Now, few interesting questions that come to mind when we think of World War-III are
\begin{itemize}
    \item Will the World War-III take place?
    \item If yes, which countries will be participating in this War?
    \item \textbf{Most Fascinating: }Which countries will be in favour and which will be against each other(inter and intra communities formed in war)?\\
\end{itemize}

\textbf{Contributions: }
In this paper we have proposed an SSN for bilateral/international relations. The edges in the network represent the relationship between nodes. Each edge has been assigned a value as score weight. These weights measure the strength of harmony or conflict (disagreement). The signed network model helps in finding groups (coalitions) that will be  involved in WWIII (World War III). The groups formed will have both less intra-group conflicting ties and less inter-group harmonious ties. To map both positive and negative relationships, a data-set of all countries in the world  has been gathered and released publicly by us. The dataset is formed by putting together various aspects that define relationships between countries such as Import and Export of goods, Border restrictions, Religious conflicts, previous wars and so on. These are discussed under \textbf{Dataset Collection and Experiments}. The \textbf{Results} Section covers the data-set analysis and prediction of  coalitions between countries fighting with each other.\\

\section{Related Works}
Social networks have been an exciting area for analysis both within and outside the scope of statistics. Modeling social models can help in developing a better understanding of network formulation and evolution.  Many social network models have come up in the past few years.  The level of density of a network can range from being densely connected to a very sparse network.  The most common network models in the literature are small world networks, scale free (or multiscale) networks, and signed network models \cite{whySSN}. Among small world networks, most general implementations include Erdos Renyi model and Watts Strogatz model. The main limitation of these models is that these models generate randomly which is not very common in real world applications. Moreover, the degree distribution in these networks vary a lot from that present in real life applications in social networks that follow the power law. To improve the applicabilty to real world problems, scale free networks come to rescue. The most familar scale free network includes Barabasi Albert model \cite{whySSN2}. They are commonly used in citation networks, world wide web, and bio-medical networks (gene regulatory networks, etc. ).  This model is best suited in situations where the network continuously grows.  To better analyze problems with a fixed number of nodes, signed social networks are best suited as they can even accommodate different types of relationships in the environment.  Other than these models, some probabilistic-based models also exist in literature \cite{whySSN1}.  Examples of such models are  Conditionally uniform distribution models, Markov graph models, and Exponential Random Graph Models. These models have application when the probability of edge formation depends upon some prior known distribution.  Thus, all models have different applicability, and the model used heavily depends on the problem under study. For the network analysis in the present work, a signed social network (SSN) will be best suited. The primary motivation behind using SSNs is because the relationships between different nations can be positive/negative. Along with this, SSN's can easily accommodate the intensity of the relationship with the help of edge weights. The use of SSN's for international relationship is also encouraged in book by Cambridge University in \cite{bookSocial}.
It is worth noting that an ample amount of ideas as well as work have been proposed using the various concepts and theorems in signed networks \cite{DBLP:journals/complexity/DuHF16} for the exposure of fruitful observations to the field of social computing. The balance theorem being the heart principle for solving discrepancies in signed networks \cite{doreian2015structural}, \cite{DBLP:conf/complexnetworks/GalimbertiMBR19},  \cite{DBLP:journals/corr/abs-1901-06845} has a lot of literature works  and has been proposed in various forms to solve different challenging problems in social networks. One such work been \cite{tzeng2020discovering} where the conflicting groups in a signed networks are discovered by involving a novel formulation in which each conflicting group is naturally characterized by the solution to the maximum discrete Rayleigh’s quotient (MAX-DRQ) problem. The approach described in \cite{tzeng2020discovering} relies on interpreting the problem objective in terms of the Laplacian of a complete graph, characterizing the spectral properties of this matrix, and deriving a novel formulation in which each conflicting
group is characterized by the solution to the maximum discrete Rayleigh quotient problem. Sometimes, balancing the whole network is not trivial for it can be better if one partitions \cite{aref2020detecting} the graph and than balance, one such idea has been implemented in \cite{doreian2009partitioning} where instead of viewing these unbalanced situations in signed networks as violations of structural balance, they can be seen as belonging to other relevant processes as call mediation, differential popularity and internal subgroup hostility. The method used in \cite{doreian2009partitioning} for partitioning the network has imposed a relaxation on the conventional balance theorem of balancing every single triangle permutation in a network. Most of existing literature pieces focuses on reducing imbalanced components in complex structural networks without considering the tolerance of these balanced networks against attacks and failures. In order to deal with such problem, \cite{wang2020robust} has taken into account the robustness of structurally balanced networks in real application for balancing processes. Here, two measures
are designed to numerically evaluate the robustness of structurally
balanced networks and different robustness performances are analyzed
to show the impact of different balancing strategies on
network robustness.

Generally, users in a structured network are connected either
positively or negatively depending on whether they agree
or disagree with another user’s opinion which leads to formation of communities across them. Several approaches have been proposed to detect these communities in unsigned networks, but not in signed network. The algorithm for community detection \cite{DBLP:journals/corr/abs-1912-07772}, \cite{DBLP:conf/icost/FangYWL16} for signed network has been proposed in \cite{anchuri2012communities} which involves an efficient two step spectral approach to
optimize modularity and other objective functions in
signed networks for community detection. Among all the curious and challenging problems in social computing, one such question as well interesting topic is the 'World War-III' and its outcomes. It is an immense topic of research in the field of social networking to predict about the groups and coalitions that will take place if World War-III occurs. Their are lot of research works going on the prediction of outcomes of World War-III and also which countries will survive after the war and ultimately emerge as superpowers. One such work is \cite{DBLP:conf/kdd/ChuWPWZC16} that has assumed a signed network where edges between vertices are either cohesive or oppositive. The method used in \cite{DBLP:conf/kdd/ChuWPWZC16} involved the novel problem of finding
k-oppositive cohesive groups from signed networks and
formulating the k-OCG detection problem as a constrained
quadratic optimization problem and designed FOCG. All the literature work described so far have only emphasised on either one problem in signed networks and have not deeply examined the features of graph for community detection. Our approach aims at prediction of coalitions among various countries in World War-III considering both the graph interior features as well as keeping in account the structural balance in the overall network formed.\\

\section{Problem Formulation}
Given a graph G consisting of nodes as countries and the edges depicting the positive and negative relationship between them,our aim is to predict the following if World War-III takes place among these countries
\begin{itemize}
    \item The best possible coalitions of countries that will be formed during war
    \item Super power countries and member countries in each communities.
\end{itemize}
\subsection{Dataset Collection}
To understand the bilateral relationships between nations, the data-set has been scrapped by us. It is formed by putting together various aspects that define relationships between countries such as, Import and Export of goods, Border restrictions, Religious conflicts, previous wars etc. The data-set covers a total of 197 nations across the globe. The data has been captured for all possible pairs counting to 38612 rows. To observe whether the relationship is one way or mutual the data is captured in both direction. In the later sections the bidirectional data is formulated to give a weight value for the relationship between all edges of complete undirected network. The weight measure the strength of harmony or conflict (disagreement). These weight will help in formulation of SSN model. The following is the complete list of aspects (factors) that have been captured to understand international relationships between nations. The list provides the description of why the consideration of particular factor is important. It also contains details of sources and range of value for factors.\\
\subsubsection{\textbf{Export and Import}}
Export refers to selling of products and services from home country to other nations across globe. Import refers to purchase of products from other countries or take services into home country. Many studies and model have found that these trade quantities directly affect bilateral relationship. \cite{10.2307/3096104} The increase in these trade quantities lead to a healthy cooperation between nations. A complete negative affect has been observed if there are conflicts in trade quantities. 

The trade value of export and imports have been web scraped using beautiful soap based python crawler from Trading Economics and Observatory of Economic Complexity (OEC). Trading Economics data is based on official sources and not any third party providers. \cite{A1} OEC on the other hand started as research project by MIT’s Collective Learning group and was later whirled out of MIT as an open source project. OEC is currently maintained by data wheel. \cite{B1} 

The value of export and imports range from 0 up to billions or trillions in USD dollars. The separate value of import and export are collected for each pair of nations. \\
\subsubsection{\textbf{Diplomatic Relation / Embassy}}
Diplomatic usually means that one is able to control a difficult situations. In context of international relations, nations usually send diplomats to work in other nation. The work of diplomats is to protect the citizens of home nation living in foreign nation, repair relationship among nations, and make long lasting bonds. Nations establish embassies in 	other nations. Embassy are the building where all work related to diplomatic missions are carried out. So study of this factor also becomes an important part while analyzing bilateral relations.

The data for study of diplomatic relation is web scraped from Wikipedia page of diplomatic mission of a nation using beautiful soap based python crawler. \cite{C1} The data is collected for all pair of nations across the globe. If the two nations involved in the pair have diplomatic relation or have embassy then they have been assigned value of 1 otherwise 0.\\

\subsubsection{\textbf{Religious Conflicts}}
“There’s no doubt that throughout history religious faith has been a major motivator for war and for destruction.” is well known claim by Richard Dawkins, famous outspoken atheist. The world around us is full of religious conflicting groups. These conflicts many a times lead to war between countries. Non Islamic nations fighting with Islamic nations is a pretty common example. Other religious conflicting groups taken into account for present study are listed in Table \ref{table:table1}. 
The data pairs of conflicting nations has been manually collected from \cite{I1}, \cite{J1}, \cite{K1}, \cite{L1}, \cite{M1}, \cite{N1}, \cite{O1}, \cite{P1}.
The list of nations has been prepared and later merged into the data-set. The value of religious conflict in the data-set is the count of religious conflicting groups between the pair of nation. The value ranges from 0 to 4. \\

\begin{table}[!h]
\small
\begin{center}
\begin{tabular}{||c c c||}
\hline
\textbf{Sr. No} & \textbf{Group 1} & \textbf{Group 2} \\ [0.5ex] 
\hline\hline
1 & Roman Catholics & Protestants\\ 
\hline
2 & Muslim-Sunni & Muslim-Shia\\
\hline
3 & Hindu & Muslim\\
\hline
4 & Jews & Roman Catholics \\
\hline
5 & Jews  & Muslim \\ 
\hline
6 & Buddhists & Muslim \\ 
\hline
7 & Christians  & Muslim \\ 
\hline
\end{tabular}
\textbf{\caption{\label{table:table1} List of Problematic Religious groupings} }
\end{center}
\end{table}

\subsubsection{\textbf{Wars}}
It is quite observant fact that the past wars between nations leads to stress on future relationship between the nations involved. Wars not only affect the both nations socially and economically. It hampers all trade activities between both nations. The wars have catastrophic effects on health and wellbeing of people in nations. \cite{murthy2006mental} This ultimately over time develops a hatred among the people of nation for the other nation affecting long term bilateral relations. The most common example of such hatred can be seen among Indians for people of Pakistan and vice versa 
The data for past wars has been captured using python crawler on Wikipedia’s list of war for each nation.\cite{D1} The data has been observed for past 50 years. The dataset contains value of 1 for presence of past war and 0 for absence of any such activity.\\
\subsubsection{\textbf{Borders}}
Border are usually lines that separate geographic areas of nations. Movement across borders for people of one nation to other can take many forms. Some nations allow complete free movement of people of bordering nation to move across, for some countries only certain group of people are allowed free movement but for other they need to take permission in form of visa from governing bodies stating the reason for movement. Some nations on other hand don’t allow people belonging to black list countries at any cost. So it is a clear observation from the types of movements, that it affects the relationship between the nations.
		
The data for borders is collected manually for all pairs of nation. The data is sourced from a variety of articles, websites and newspaper ranging from Wikipedia articles to here \cite{E1}. The values assigned to each pair is picked from vector $<$ -1, 0, 1, 2 $>$. The Table \ref{table:table2} Provides the map of each value to the type of movement. \\

\begin{table}[!h]
\small
\begin{center}
\begin{tabular}{||c c||}
\hline
\textbf{Value} & \textbf{Description}\\ [0.5ex] 
\hline\hline
-1 & Completely closed Border\\ 
\hline
+2 & Completely open border for both nationals\\
\hline
+1 & Open border but no legal agreement \\
\hline
0 & No agreement as such (Need Visa to visit)\\
\hline
\end{tabular}
\textbf{\caption{\label{table:table2}Description for 'Border-Values' used in the database} }
\end{center}
\end{table}

\subsubsection{\textbf{Peace Treaties}}
Peace treaties is usually legal agreement that is done between conflicting groups to end the hostility. Nations usually sign peace treaties to formally end state of war between them. This is usually followed by settlement of conflict and start of trade to promote harmony. Thus peace treaty act as a treaty of friendship, co-operation and stabilization of relation. 
The data for peace treaties is collected manually for all pairs from sources ranging from Wikipedia list of treaties to \cite{F1}. The data-set contains value of 1 if peace treaty is signed between nation and 0 otherwise.\\
\subsubsection{\textbf{Exchange Rate Ratio} (Foreign Direct Investments)}
Exchange rate refers to value of currency from one nation to other. Exchange rate is determined by study of open market supply and demand of currency and also by ongoing trade activities. Exchange rate ratio is defined as ratio of exchange rate of nation A to exchange rate of nation B on same currency scale.

\begin{equation}\nonumber\label{eq1}\scriptsize
\begin{split}
Exchange\ Rate\ Ratio &= \frac{Exchange\ Rate\ of\ Nation\ A\ (in\ EUR)} {Exchange\ Rate\ of\ Nation\ B\ (in\ EUR)}
\end{split}
\end{equation}

If the exchange rate ratio for nation A to B greater than 1, then A’s investors will be benefited by investing in nation B. On similar notes if exchange ratio is less than 1, then A’s investors will not be benefited as profits will get cut due to exchange rate involved. So this ratio has direct effect on foreign direct investments in a nation. More the ratio more the positive relation and vice versa. 
In data-set exchange rate of nations have been calculated in euro (EUR). The data is collected using API developed by fixer.io. Fixer is an open source API for current and historical foreign exchange rates that are published by European central bank. \cite{G1} The ISO codes required for API was scrapped using pycountry module of python. 

The value of exchange rate ratio is a real valued number. \\
\subsubsection{\textbf{International Cases}}
International court of justice (ICJ) was founded in 1945. It is situated in Netherlands. The main work of ICJ involves settling conflicts between nations in accordance to international laws. The disputes usually are related to border issues, terrorism, people captured on line of controls etc. So international cases help in analysis of relationship between nations.

The data has been collected manually from official records of cases available on ICJ 	website. \cite{H1} The data contains value of 1 for pair which has a case registered in ICJ and 0 otherwise. The cases collected for data-set range from 1950s to 2020.

\begin{table}[H]
\begin{minipage}{0.5\textwidth}
\begin{center}
\scriptsize
\begin{tabular}{||p{0.2cm} p{2.1cm} p{1.35cm} p{1.9cm} p{0.8cm}||}
\hline
\textbf{Sr. No} & \textbf{Factor} & \textbf{Value Range} & \textbf{Reference Name} & \textbf{Scraping Category}\\ [0.5ex] 
\hline\hline
1 & Exports & 0 to trillion dollar & Trading Economics, OEC & Web crawler\\ 
\hline
2 & Imports & 0 to trillion dollar & Trading Economics, OEC & Web crawler\\
\hline
3 & Religious conflicts & {0, 1, 2, 3, 4} & Wikipedia & Manual\\
\hline
4 & Diplomatic relations/ Embassy & {0, 1} & Wikipedia & Web crawler \\
\hline
5 & Past wars  & {0, 1 } & Wikipedia & Web crawler \\ 
\hline
6 & Border movement  & {-1, 0, 1, 2 } & Wikipedia & Manual \\ 
\hline
7 & International cases  & {0, 1 } & ICJ official site & Manual \\ 
\hline
8 & Peace treaties  & {0, 1 } & Wikipedia & Manual \\ 
\hline
9 & Exchange rate ratio & Real number & Fixer.api & Python script \\ 
\hline
\end{tabular}
\textbf{\caption{\label{table:table3}Short description of all factors} }
\end{center}
\end{minipage}
\end{table}

\subsection{Dataset Normalization }
The data-set contains factors that map to different set of values. Factors like export and import, exchange rate has very wide range of values, on contrary factors like religious conflict, treaties, war, border etc. have value ranging over small set. Therefore there is need to normalize the data 	into a common value domain for easy calculation of score. Score can be later used for evaluation of strength and type of tie.
The normalization of i-th row for factor ‘y’ (other than border and exchange rate ratio) of data-set been carried out by use of formula given below:
\begin{equation}\label{eq2}
  \begin{split}
y_{n}(i) & = \frac{y(i)-Min(y)}{Max(y)-Min(y)} \\
\end{split}  
\end{equation}
The value for border factor has been normalized by the formula given below:
\begin{equation}\label{eq3}
  \begin{split}
y_{n}(i) & = \frac{y(i)}{Max(y)} \\
\end{split}  
\end{equation}
where,\\
$y_{n}(i)$\hspace{0.2cm}-\hspace{0.2cm}normalised value of factor y at $i^{th}$  index of data-set\\
y(i)\hspace{0.2cm}-\hspace{0.2cm}normalised value of factor y at $i^{th}$  index of data-set\\
I\hspace{0.2cm}-\hspace{0.2cm}set of all possible index values for rows in data-set\\
Min(y)\hspace{0.2cm}-\hspace{0.2cm}minimum value of factor y over all $i\in I$\\ 
Max(y)\hspace{0.2cm}-\hspace{0.2cm}maximum value of factor y over all $i\in I$\\

The value for exchange rate ratio has been normalized using function $f(y_{i}):\mathbb{R} \to \{-1,1\}$ given by

\begin{equation}\label{eq4}
f(y_{i})=
\begin{cases}
1 & \text{if} \hspace{1cm} y_{i} \geq 0  \\    
-1 & \text{if} \hspace{1cm} y_{i} <  0  \\  
\end{cases}
\end{equation}
where $y_{i}$ is value of factor y at $i^{th}$ index.

The normalization of all factors using above methods result in range of values to lie between -1 to 1. The normalized data-set is exported and saved externally in comma separated value (csv) format for further study.

\subsection{Score Calculation}
The normalized dataset is used for score calculation for each pair of nations. These values will help in determining the strength of harmony or conflict (disagreement). It will also assist in the development of signed model network. First, we gather all the factors that can be used to model the relationship between two nations. These factors are already discussed in Table \ref{table:table3}. To understand the weightage of each factor, we assign coefficients to each of them. These coefficients will help the model better analyze the weighted importance in the final score calculation. But the coefficients need to be hyper-tuned before use in the designing of the model, and one needs some evaluation set for comparison of different parameter values and find the best coefficients for the model. We developed an evaluation set of the most commonly known relationships among nations for cross-validation to overcome these issues. The evaluation set cannot associate a weight value with itself; instead, it stores the relationship type ( positive/ negative). Also, we try to model the complex relationship using a linear equation in terms of factor value and its coefficient. We expect that such relations seem too complicated but can be estimated well enough using linear models. Coming up with nonlinear functions can be an interesting open direction. The linear function used in our model is given by Eq. \ref{eq:score}. To begin the hyper tuning, we initialize all the coefficients to have an equal weightage of 0.125 (=1/8). We find the score values for all pairs in the evaluation set and compute our model's accuracy with these coefficients by comparing the signs of the final score calculated. To improve upon the model's efficiency, we manually fine-tune different coefficients by re-weighting the coefficient based upon the role each plays in valuing the relationship. The best-found hyper-parameters after an intensive search are reported in Table \ref{table:table4}.

The score value is a real value which can be either negative, positive or zero. The formula used for evaluation is given below:
\begin{equation}
\begin{split}
\textbf{Score}(y_i) = e* y_{i}(export) + i * y_{i}(import) +\\ 
b*y_{i}(border) + d* y_{i}(diplomatic) + p* y_{i}(treaty)\\ 
+ x*y_{i}(exchange\ rate\ ratio) - w *y_{i}(war)\\ 
- c*y_{i}(int.cases) - r*y_{i}(religious\ conflict)
\end{split}
\label{eq:score}
\end{equation}
where e,i,b,d,p,x,w,c and r are  factor coefficients  and  $y_{i}(P)$  is  value of factor y at  $i^{th}$  index for factor P
After the value of $S(y_{i})$ is calculated for each pairs of nations across both direction, these are then combined to have single undirected score value between nations for signed network model. 
This will help us in understanding overall mutual relationship. 
This result in collection of 19306 rows in data-set for 197 nations across globe.

After Normalization and Score calculation, the dataset is ready for use in any real world application involving bilateral relation between nations. 

\begin{table}[H]
\scriptsize
\begin{center}
\begin{tabular}{||c c c c||}

\hline
\textbf{Sr. No} & \textbf{Factor} & \textbf{Symbol} & \textbf{Coefficient Value}\\ [0.5ex] 
\hline\hline
1 & Exports & e & 5.0\\ 
\hline
2 & Imports & i & 5.0\\ 
\hline
3 & Religious Conflicts & r & 2.0\\ 
\hline
4 & Diplomatic relations / Embassy & d & 0.8\\ 
\hline
5 & Past Wars & w & 3.0\\ 
\hline
6 & Border movement & b & 2.0\\ 
\hline
7 & International cases & c & 0.5\\ 
\hline
8 & Peace treaties & p & 0.125\\ 
\hline
9 & Exchange rate ratio & x & 0.5\\ 
\hline
\end{tabular}
\textbf{\caption{\label{table:table4}Coefficient for bilateral relationship} }
\end{center}
\end{table}

\subsection{Proposed Algorithm}\label{algo}
In previous sections we have calculated the normalized scores for all country-pairs, which will help in formulation of a SSN model. The developed SSN model will help in finding groups (gangs) that will be involved in WWIII (World War III). The groups formed will have both less intra-group conflicting ties and  inter-group harmonious ties. For developement of SSN model we use a networkx graph with countries as nodes and scores as edge attribute ‘weight’.  The initial graph formed will be highly unstable. As seen in Figure 1, if any triangle with 3 countries has edge weights in the order of (negative,negative,negative) or (positive,positive,negative), they are termed as unstable-triangles. These unstable triangles have tendency to get converted to (positive,negative,negative) or (positive,positive,positive) stable states.\\
The generalized algorithm for the conversion of unstable to stable triangles is highly randomized as it selects a random unstable triangle and toggles any one edge randomly. This process cannot be applied to our SSN model as the dataset is large having total of 1254890 triangles. Also the generalized algorithm doesn't guarantee a definite decrease in number of unstable triangles. Moreover in our case just random toggling  of a negative relation as positive or vice versa cannot be justified.\\
So, we used a modified version of generalized algorithm as described by Algorithm \ref{algorithm_bal_modified}, the procedure unstable-to-stable when called with a list of all triangles and their corresponding weights, converts one triangle from unstable to stable state. The algorithm chooses an unstable triangle randomly, but the edge to toggle is chosen as the one with the least magnitude of the 3. Also, the weight of the new edge if changing from ‘-’ to ‘+’ is updated as the sum of the magnitudes of the remaining edges. This is done keeping in mind that the very reason for such a change in relation is due to the mutual enmity with a common enemy, so the score is accordingly the sum of their enmity. Similarly, if the edge is changing from ‘+’ to ‘-’, its weight is updated as sum of the remaining edges- its previous weight. This is done to make sure that a sudden new huge enmity is not introduced into the dataset.\\
The procedure unstable-stable can be called after re-evaluating the number of unstable-triangles, infinitely until all the triangles are stabilized(theoretically). Practically, in real world applications it is observed that there is a threshold
\begin{algorithm}[H]
\caption{Our Modified Algorithm for Structural Balance in triangles}
\label{algorithm_bal_modified}
\begin{algorithmic}[1]
\Procedure{unstable-to-stable}{G, tris-list, all-weights}
\While{unstable-triangle not found}
\State Generate a random index
\State Check if the corresponding triangle in all-weights is unstable
\EndWhile 
\State least-edge = find edge with least abs(Wt)
\If{least-edge's sign = -}
\State Change it to +
\State Update Wt as sum(abs(Wt of remaining 2 edges)
\ElsIf{least-edge's sign = +}
\State Change it to -
\State Update Wt as sum(abs(Wt of remaining 2 edges)-it's prev Wt)
\EndIf
\State return G
\EndProcedure
\end{algorithmic}
\end{algorithm}
\noindent
value after which number of stable triangles is almost constant (i.e. not decreasing to 0 and fluctuates within the same range). In the present work the threshold is taken to be 5 lakh unstable-triangles left, out of 12.5 lakhs total triangles.\\
Once the graph is near-stable, we can try and predict the coalitions to-be formed. The generalized algorithm for the same starts by placing a random node in the first list and recursively visiting all its neighbors placing nodes with positive relation in the same list and negative ones in the other list. The generalized algorithm assumes the graph to be completely stable, which isn't true in our case. So,  an improvised Algorithm \ref{algorithm_form_coalition}  is proposed.\\
\begin{algorithm}[H]
\caption{Our Modified Algorithm for forming coalitions}
\label{algorithm_form_coalition}
\begin{algorithmic}[1]

\Procedure{single-append}{G, country, List1, List2}   
\While{unprocessed neighbours of country}  
        \State Find the strongest positive neighbour
        \State Find the strongest negative neighbour
    \EndWhile 
\If{country in List-x} \Comment{x = 1 or 2}
        \State Add positive neighbour in list x
        \State Add negative neighbour in list y
    \EndIf
    \State return List1,List2,positive and negative neighbour
\EndProcedure
\Procedure{see-coalitions}{G,start-country}
\State Append start-country in L1 and to-check-pos
\While{to-check-pos or to-check-neg}
    \State L1,L2,pos,neg=single-append(G,to-check-pos[0],L1,L2)
    \State Append pos,neg to to-check-pos,to-check-neg 
    \State delete to-check-pos[0]
    \State L1,L2,pos,neg=single-append(G,to-check-neg[0],L1,L2)
    \State Append pos,neg to to-check-neg,to-check-pos 
    \State delete to-check-neg[0]
    \EndWhile
    \State return L1,L2

\EndProcedure
\end{algorithmic}
\end{algorithm}
\begin{minipage}{0.97\textwidth}
\begin{minipage}{0.48\textwidth}
The procedure “single-append” when given a nation, appends its most-positive relations neighbor to list 1 and its most-negative relations neighbour to list 2.\\
While the procedure “see-coalitions” when given a starting nation, appends it to the list 1 and calls “single-append” procedure on this country. It then continues similarly for the returned nations. So, every nation starting from the input nation(also including those of the second list unlike the general algorithm), gets to choose its strongest ally and its greatest enemy. This situation is exactly what we see in world wars, as there is no global reason but a collection of local reasons to escalate the fight.\\
As the algorithm  depends largely on the first nation chosen, an automated script has been designed. The script
\end{minipage}
\hfill
\begin{minipage}{0.48\textwidth}
calls the above described functions on all nations as the first node and the result is validated with another dataset of 43 such well known Allies and Enemies pairs (manually prepared).
The final results obtained will be analysed  and discussed in the results section of this paper. The code, dataset and all figures are available publicly\footnote{https://github.com/worldWarIII-scomp/WorldWarIII\_Scomp}.
\section{Results}
\subsection{Data-set Collection findings}
In the present study data-set comprising of 38612 rows is collected. The prepared data-set is visualised in Figure xx against all aspects. This will provide a better understanding of diversity of data-set.
\end{minipage}
\newpage
\vspace{1.5cm}
\begin{minipage}{0.97\textwidth}
\begin{figure}[H]
    \centering
    \hfill
    \begin{subfigure}[b]{0.47\textwidth}
         \centering
         \includegraphics[width=\textwidth]{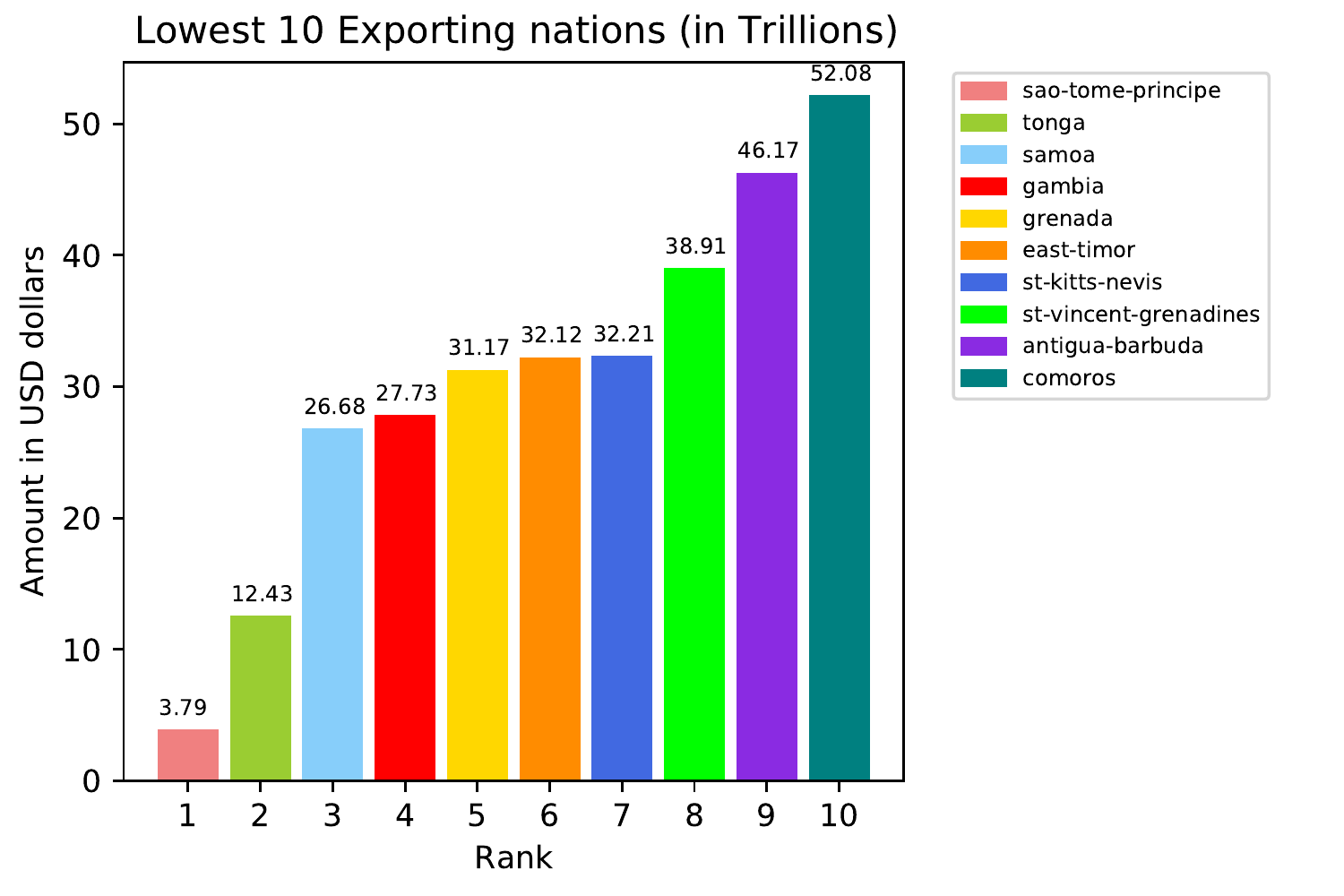}
         \label{subfig:1}
     \end{subfigure}
     \hfill
     \begin{subfigure}[b]{0.47\textwidth}
         \centering
         \includegraphics[width=\textwidth]{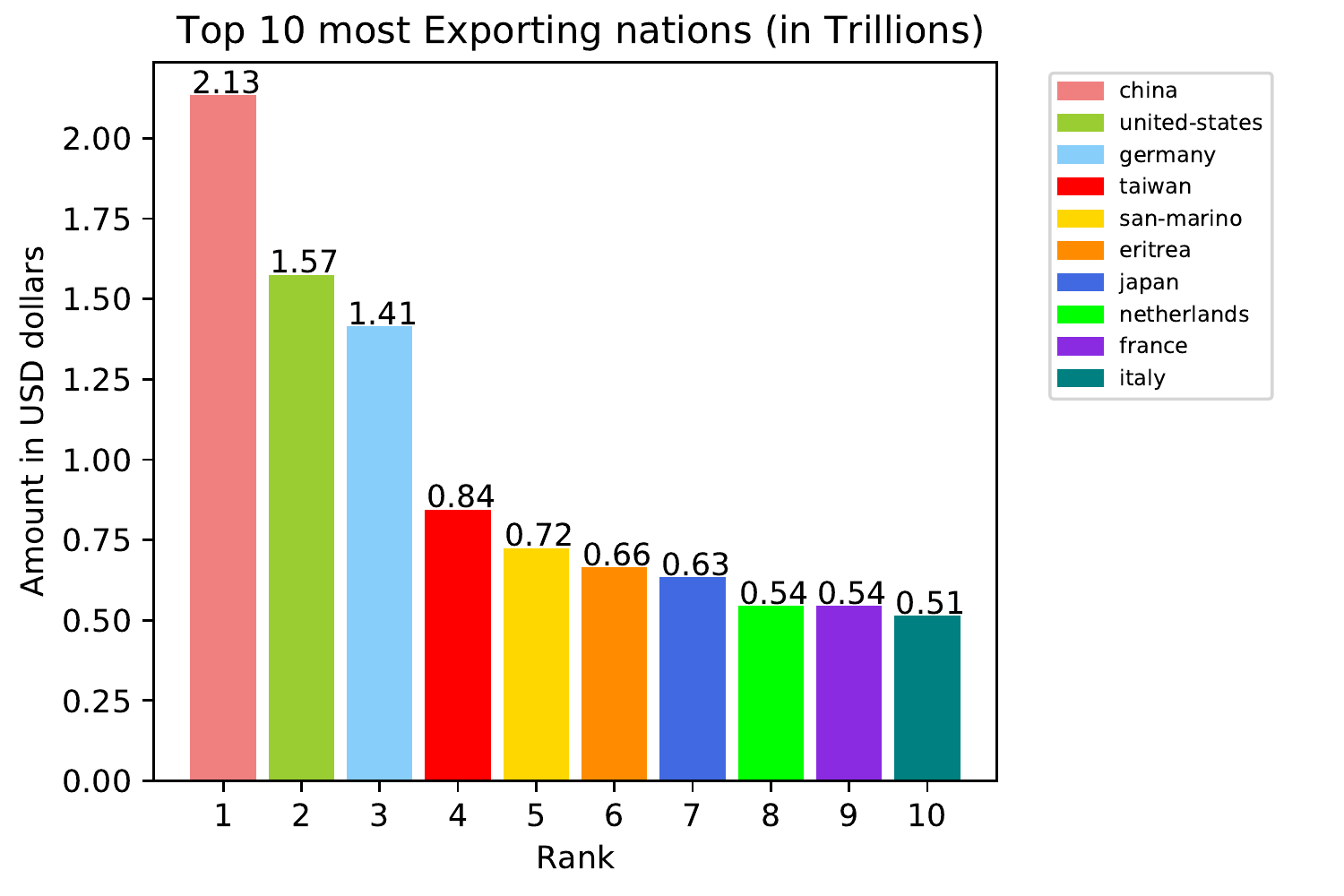}
         \label{subfig:2}
     \end{subfigure}
     \hfill
     \begin{subfigure}[b]{0.47\textwidth}
         \centering
         \includegraphics[width=\textwidth]{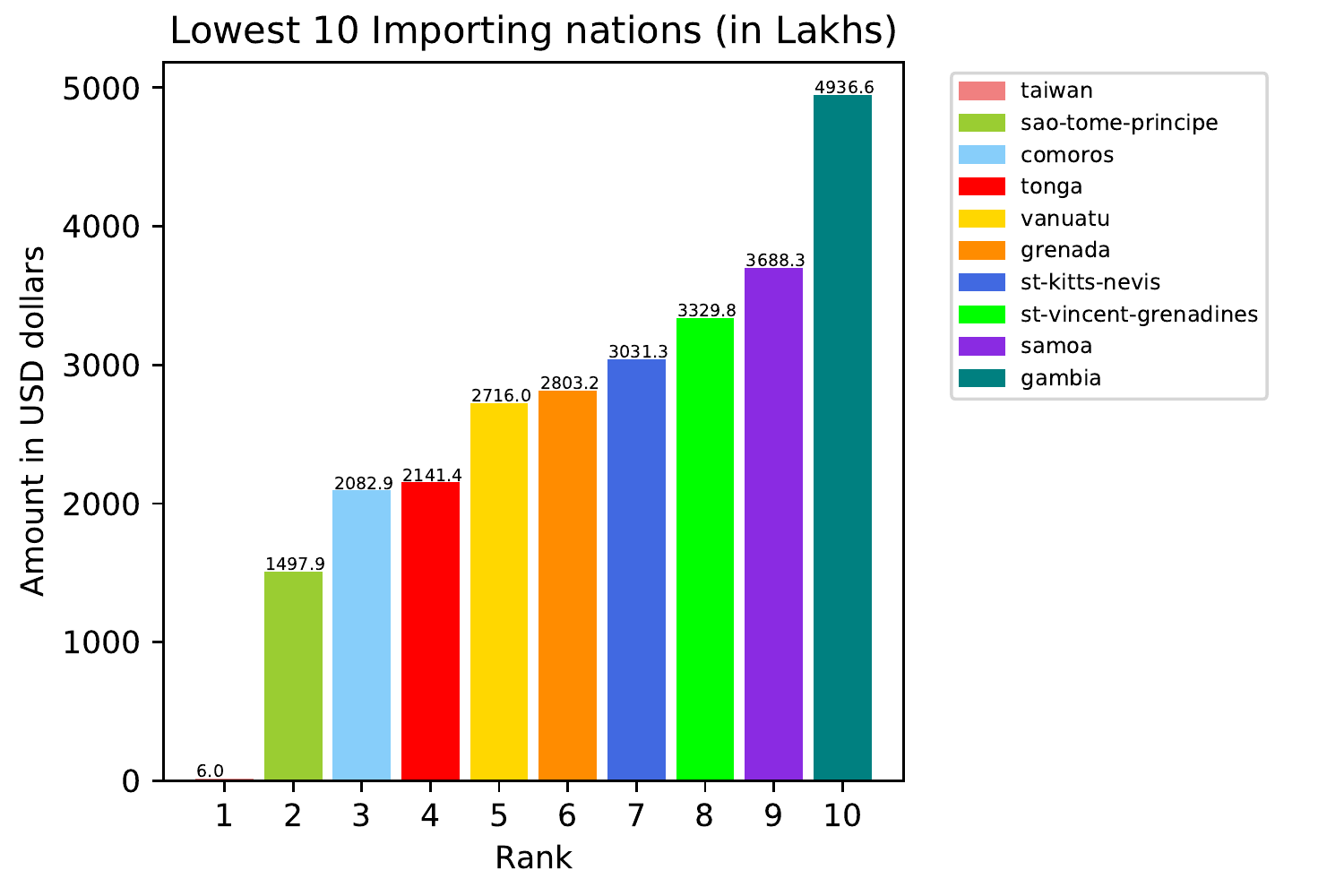}
         \label{subfig:3}
     \end{subfigure}
     \hfill
     \begin{subfigure}[b]{0.47\textwidth}
         \centering
         \includegraphics[width=\textwidth]{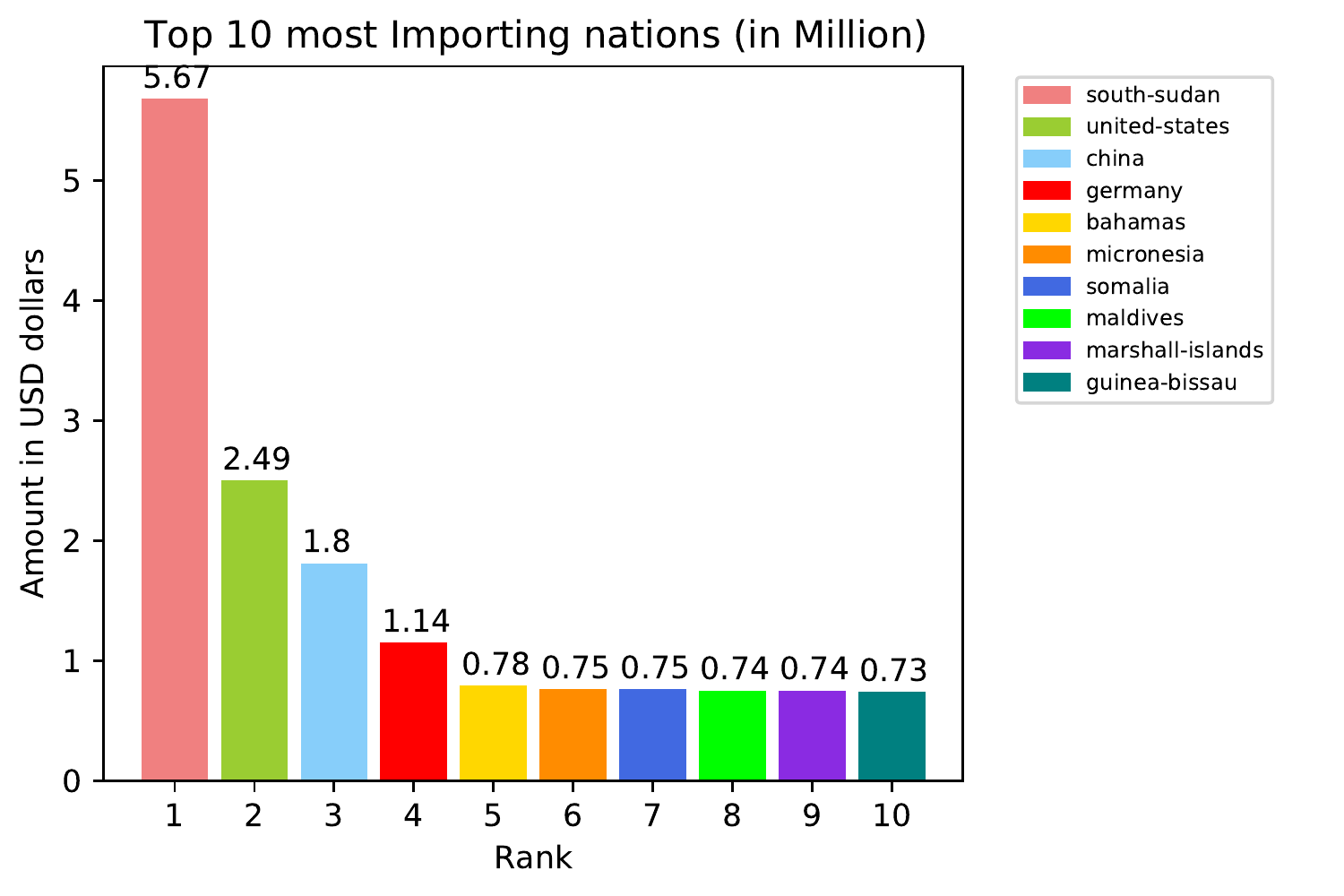}
         \label{subfig:4}
     \end{subfigure}
    \caption{Plots for visualization of dataset based on different factors can play role in formation of signed network.}
    \label{Complete_figure}
\end{figure}
\end{minipage}
\end{minipage}
    \begin{figure*}[h]\ContinuedFloat
     \centering
     \hfill
     \begin{subfigure}[b]{0.47\textwidth}
         \centering
         \includegraphics[width=\textwidth]{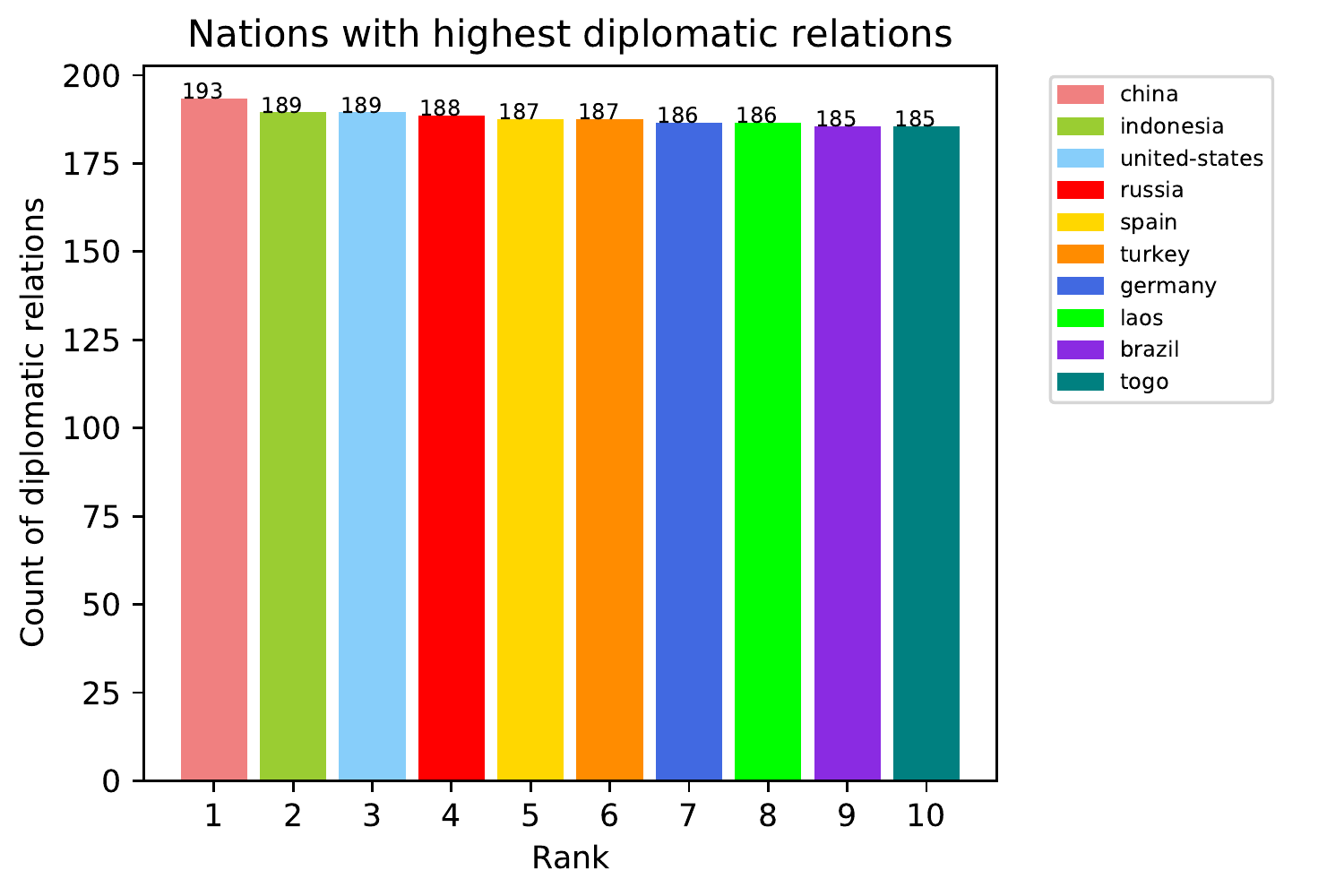}
         \label{subfig:5}
    \end{subfigure}
    \hfill
     \begin{subfigure}[b]{0.47\textwidth}
         \centering
         \includegraphics[width=\textwidth]{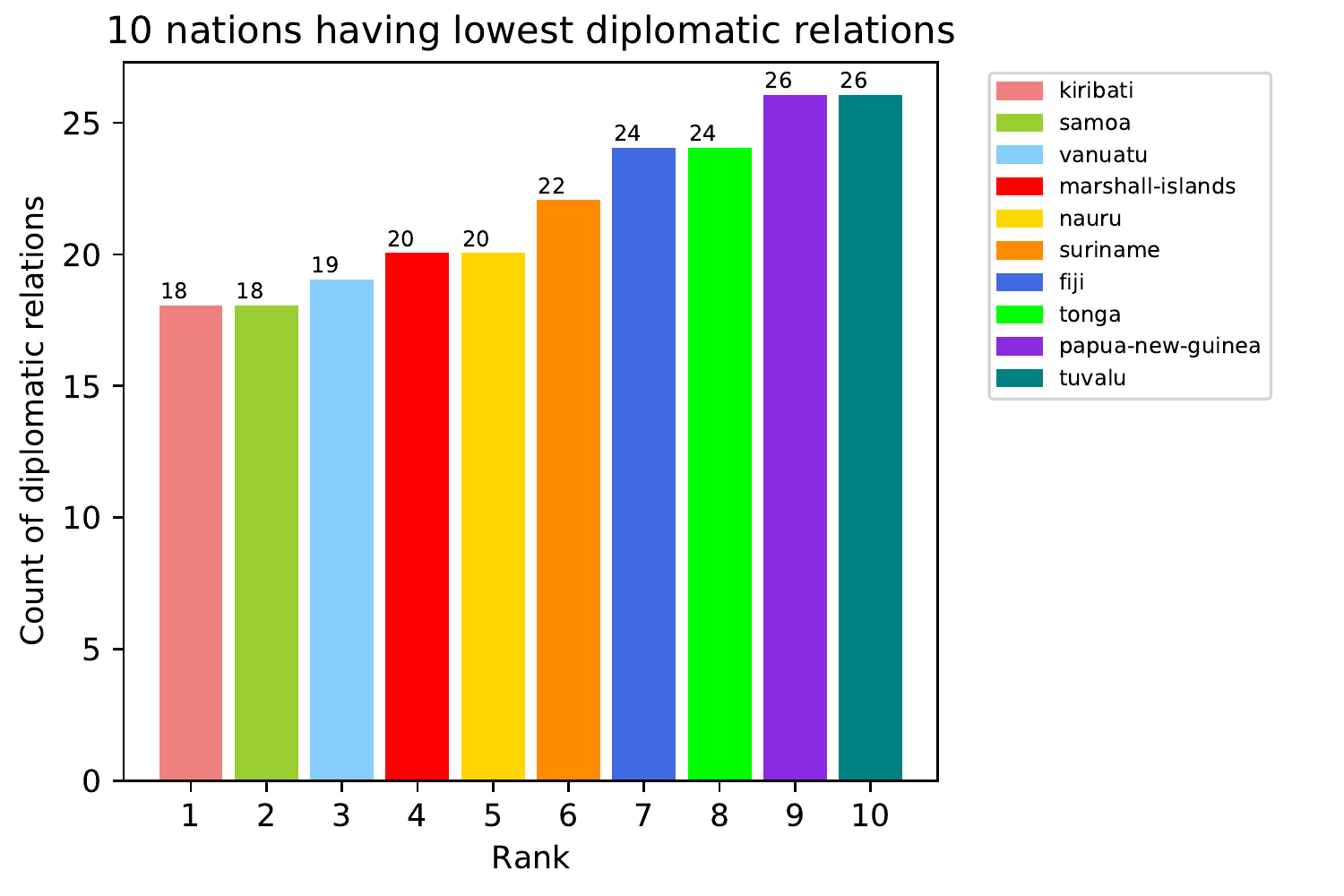}
         \label{subfig:6}
     \end{subfigure}
     \hfill
     \begin{subfigure}[b]{0.47\textwidth}
         \centering
         \includegraphics[width=\textwidth]{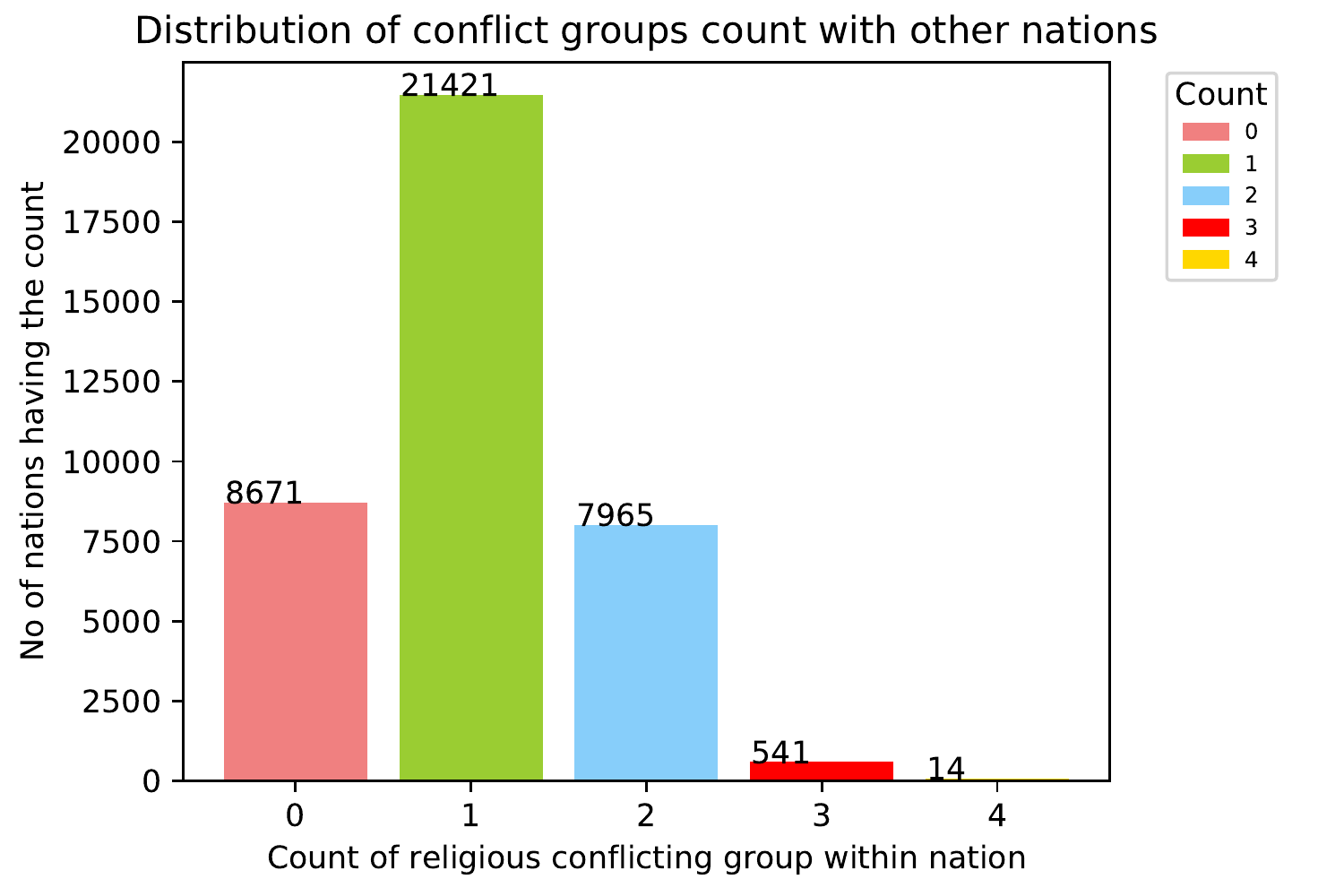}
         \label{subfig:7}
     \end{subfigure}
     \hfill
     \begin{subfigure}[b]{0.47\textwidth}
         \centering
         \includegraphics[width=\textwidth]{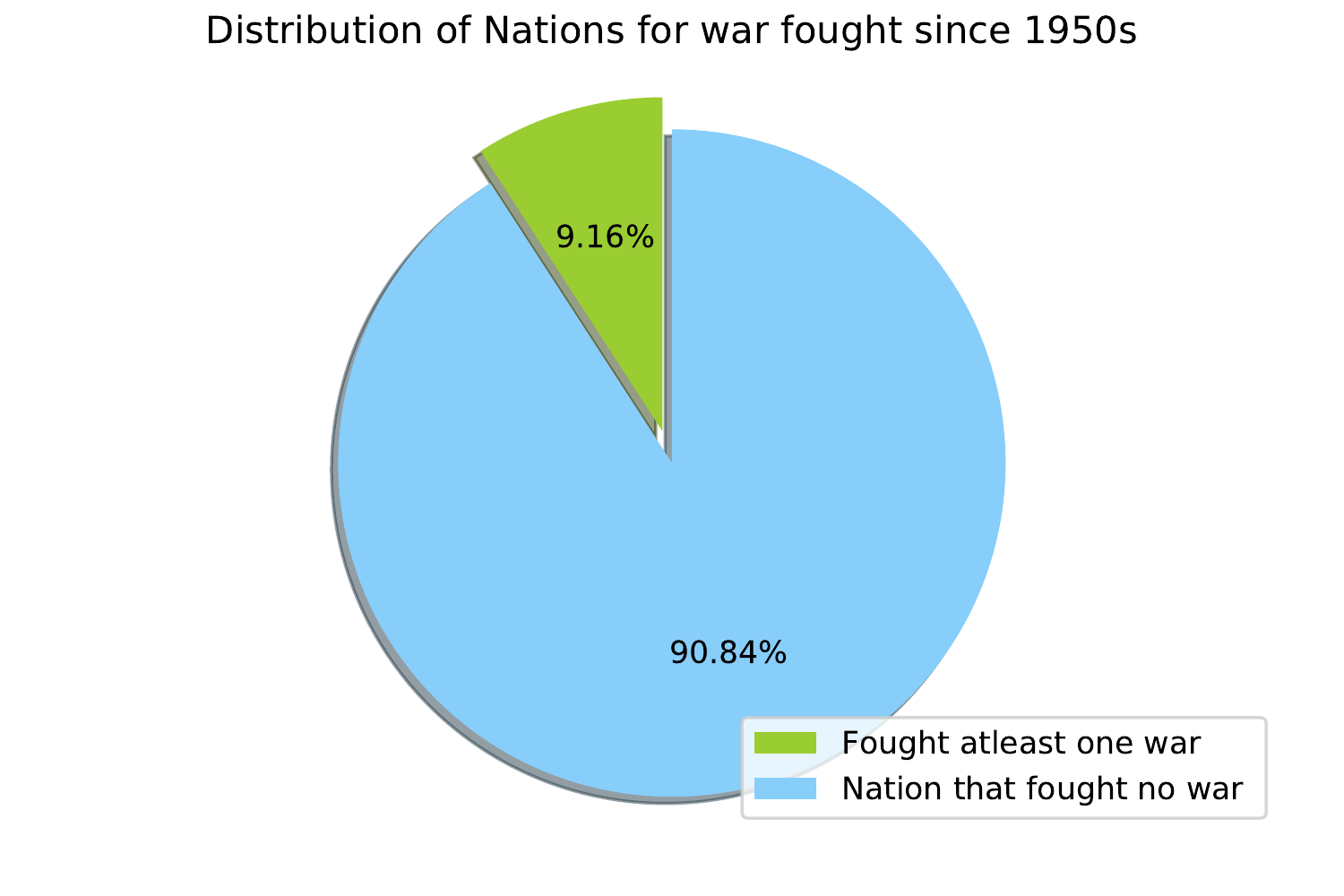}
         \label{subfig:8}
     \end{subfigure}
     \hfill
     \begin{subfigure}[b]{0.47\textwidth}
         \centering
         \includegraphics[width=\textwidth]{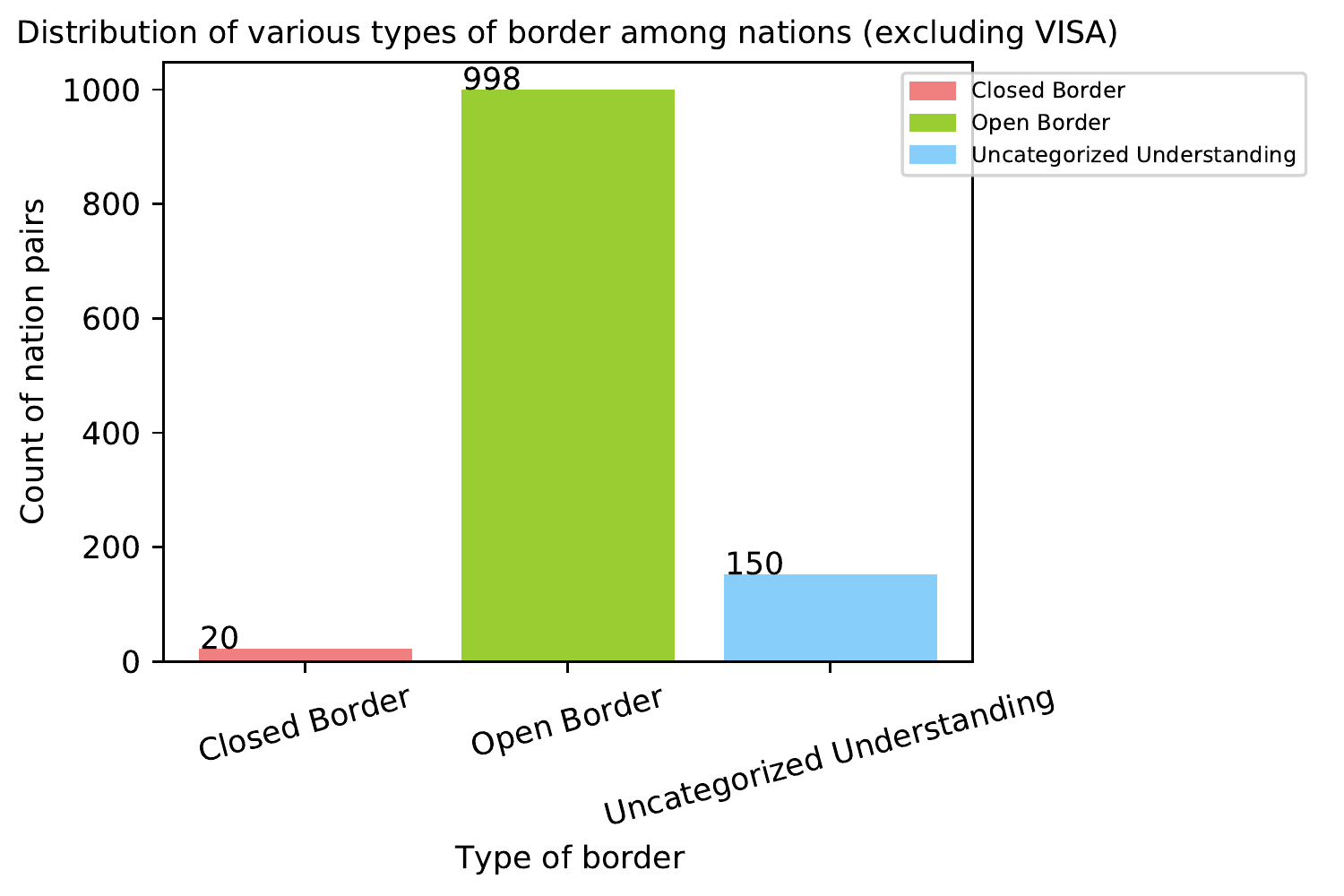}
         \label{subfig:9}
     \end{subfigure}
     \hfill
     \begin{subfigure}[b]{0.47\textwidth}
         \centering
         \includegraphics[width=\textwidth]{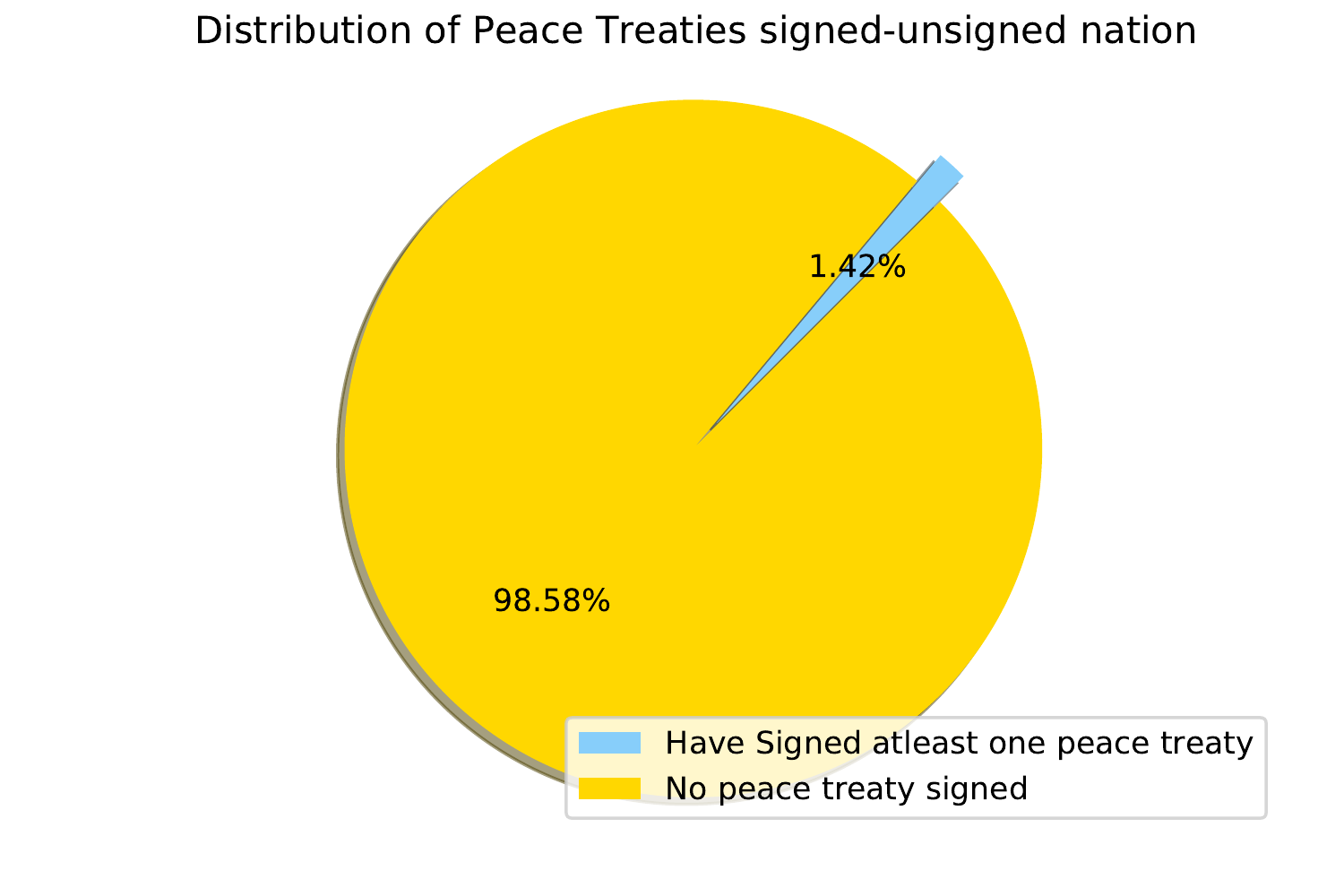}
         \label{subfig:10}
     \end{subfigure}
     \hfill
     \begin{subfigure}[b]{0.47\textwidth}
         \centering
         \includegraphics[width=\textwidth]{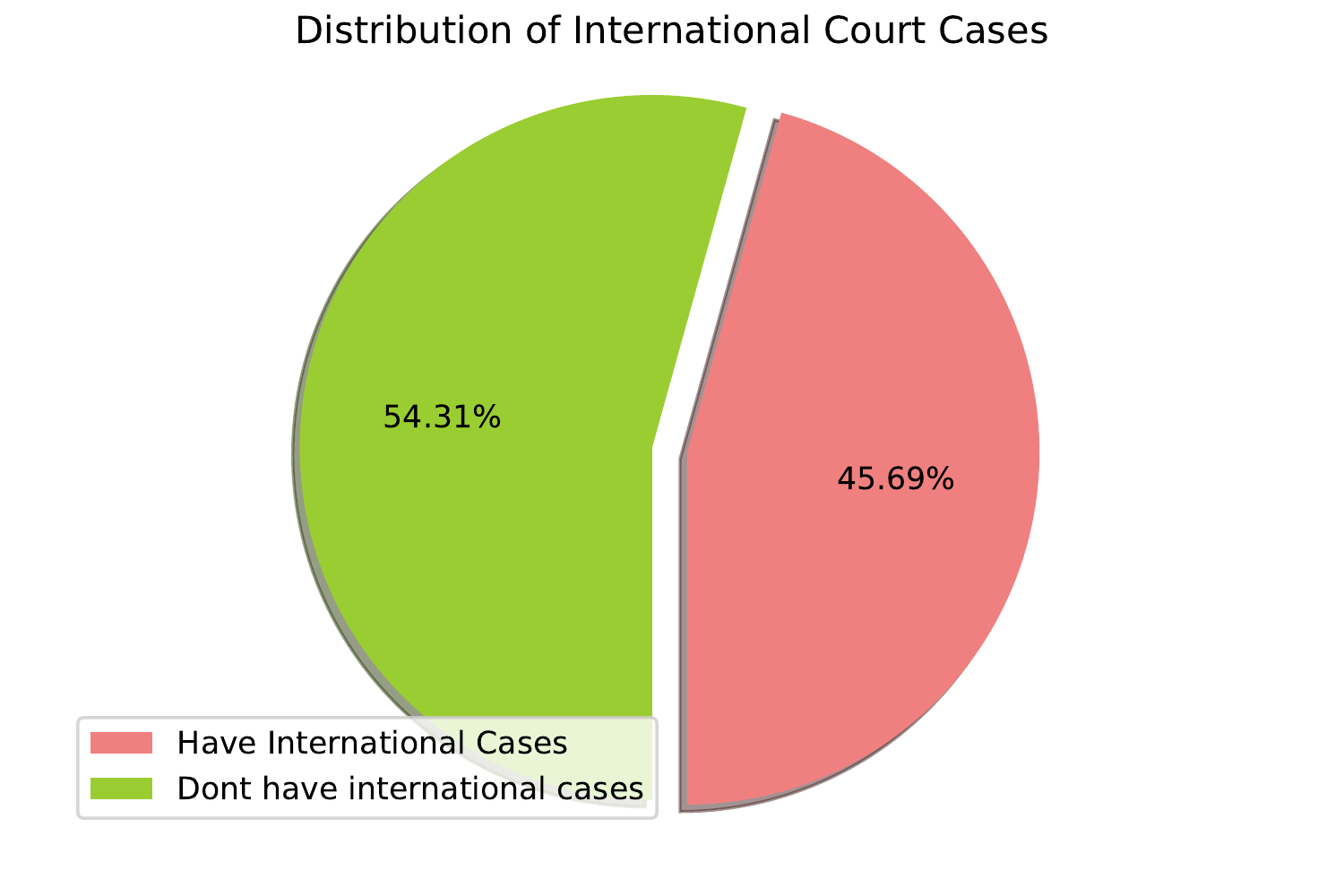}
         \label{subfig:11}
     \end{subfigure}
        \caption{(Contd.)Plots for visualization of dataset based on different factors can play role in formation of signed network.}
        \label{Complete_figureContd}
\end{figure*}
\cleardoublepage
\textbf{\color{blue}Figure 2 Interpretation:}
\begin{itemize}
    \item Exports \& Imports are considered an important factor in bilateral relations since it benefits both nations involved with resources of each other. From the graphs, China is the highest exporting country and Taiwan is the lowest importing country.
    \item Each nation across the globe has a certain number of conflicting religious groups. These groups tend to have negative relations with groups present in foreign nations. It can be seen that almost a large number of nation pairs have at least one religious conflicting group.
    \item It can be clearly seen that only about 9.16 percent of nations' pairs have been involved in wars. This signifies that nations usually avoid involving in war as it has catastrophic effects on nations growth.
    \item It can be observed that many nations tend to make the movement across borders open and free for all populations  or some restricted set of populations in both nations. This is done in order to make trade and bilateral relations harmonious.
\end{itemize}
\subsection{Prediction Experiment findings}
\vspace{2.5mm}
\subsubsection{Signed network and Coalitions}
With the use of modified algorithms for structural balancing and coalition formation discussed in section \ref{algo}. The two coalitions so formed of 197 nations is shown in Table \ref{table:table5}. The table contains both sets of coalitions that will probably be fighting opposite each other in world war III.\\
The visual plots of these coalitions are shown in Figures 3 \& 4.
In Figure 3, we tried to plot all the 197 countries into these coalitions with yellow colored nodes belonging to coalition set I and white nodes belonging to set II. The positive and negative edge relationships are represented by blue and red color respectively. Since this graph was quite dense, we plot a sub-graph version with key countries, and is shown in Figure 4; where yellow nodes belong to set I, green ones to set II and blue \& red edges represent positive and negative relations respectively.\\
\textbf{\color{blue}Interpretation:}
\begin{itemize}
\item From Table \ref{table:table5},  we observe that out of 197 nations  99 belong to coalition set I and 98 to coalition set II. So one can see that world war III will not contain unbalanced coalition sets in terms of count of nations on each side.
\item It can be seen in  Figure 4 that most of the edges within both coalition sets are blue (positive) and edges across both sets are red(negative) which is in correspondence with the balance theorem described in ourpaper. However, one can observe that not all intra (coresp. inter) edges are positive (coresp. negative) reflecting that this is due to the reason for our signed network is balanced upto a ratio of 0.565 ie. (709690/1254890 triangles). This ratio acts as the threshold value in our work as it cannot be further improved within computational limits.
\item Even at the threshold value one can note that the model predicts well known pairs of allies and enemies on the expected sides of coalitions. For instance pairs like USA - China , USA - North-Korea, India -Pakistan, are predicted to be on opposite sides of coalitions. Similarly pairs like USA-India , Turkey - Azerbaijan that have good bilateral relations are observed to be on the same side during World War III. We will discuss the test set of such well known relations in the next sub-section and one can spot all these countries easily in the subgraph in Figure 4.
\item Only a few pairs like North-Korea - South-Korea  are observed on the same side which is not as per one's expectation. This is due to the fact that one cannot achieve complete 100\% of stable triangles hence leaves scope for some intra negative ties within coalition sets. 
\end{itemize}
\begin{minipage}{0.87\linewidth}
\begin{table}[H]
\small
\begin{center}
\begin{tabular}{||c | c||}
\hline
Set 1 & Set 2\\ [0.5ex] 
\hline\hline
\hspace{0.5mm}
\begin{minipage}{0.47\linewidth}
\vspace{1mm}
  Austria, Egypt, Luxembourg, Azerbaijan, Equatorial-Guinea, Australia, Lebanon, Brunei, Slovenia, Oman, Liechtenstein, Ethiopia, Malaysia, France, Chad, Japan, Djibouti, Lithuania, Sierra-Leone, Iceland, South-Africa, Poland, Costa-Rica, Benin, South-Sudan, Turkey, Hungary, Switzerland, India, Myanmar, East-Timor, New-Zealand, Portugal, Niger, Ireland, Spain, Turkmenistan, Grenada, Malta, Finland, Canada, Zimbabwe, Monaco, Kyrgyzstan, Tanzania, Sweden, vatican-city-state, United-Arab-Emirates, Taiwan, Singapore, st-kitts-nevis, Belgium, Kenya, Panama, Qatar, Estonia, Rwanda, Venezuela, Ukraine, Guinea, Afghanistan, Norway, bosnia-herzegovina, Kuwait, United-States, Micronesia, Yemen, Angola, Guinea-Bissau, Marshall-Islands, Albania, Algeria, Fiji, San-Marino, Bolivia, Saint-Lucia, Burkina-Faso, Samoa, Solomon-Islands, Latvia, Tuvalu, Barbados, Vanuatu, Chile, Germany, Zambia, antigua-barbuda, Nepal, Czechia, Bulgaria, Mauritania, Togo, Peru, Nauru, Palau, Sri-Lanka, Maldives, Papua-New-Guinea, Mongolia
\end{minipage}

\hspace{0.5mm}
&
\hspace{0.5mm}
\begin{minipage}{0.46\linewidth}
\vspace{1mm}
\begin{tabular}{p{\linewidth}}
  Haiti, st-vincent-grenadines, Syria, Jamaica, Madagascar, Iran, Pakistan, Guyana, Uzbekistan, dominican-republic, United-Kingdom, South-Korea, Eswatini, Liberia, China, Kosovo, El-Salvador, Kazakhstan, macedonia, Iraq, Nigeria, Comoros, Tunisia, Indonesia, Philippines, Mali, Trinidad-Tobago, Thailand, Tajikistan, Jordan, Serbia, Mozambique, Bhutan, Romania, Armenia, Burundi, Gabon, Denmark, Tonga, Libya, Saudi-Arabia, Kiribati, Nicaragua, Gambia, Russia, Belize, Guatemala, Colombia, Georgia, congo, Bahrain, Senegal, Eritrea, Suriname, Seychelles, Mexico, Palestine, republic-congo, Sudan, Morocco, Botswana, Croatia, Brazil, Lesotho, Honduras, Mauritius, Ivory-Coast, Slovakia, Greece, Uruguay, Cuba, Andorra, Cameroon, Ghana, Uganda, Netherlands, Israel, Paraguay, Bangladesh, Dominica, Malawi, Moldova, Ecuador, Italy, central-african-republic, Namibia, Argentina, Cambodia, Belarus, North-Korea, sao-tome-principe, Montenegro, Somalia, Bahamas, Cape-Verde, Laos, Vietnam, Cyprus\\
\end{tabular}
\vspace{1.5mm}
\end{minipage}
\hspace{0.5mm}
\\
\hline
\end{tabular}
\textbf{\caption{\label{table:table5}Coalitions of Nations predicted by our SSN model. } }
\end{center}
\end{table}
\end{minipage}


\begin{figure*}[hpt!]
    \begin{center}
    \includegraphics[width=18cm,height=22cm]{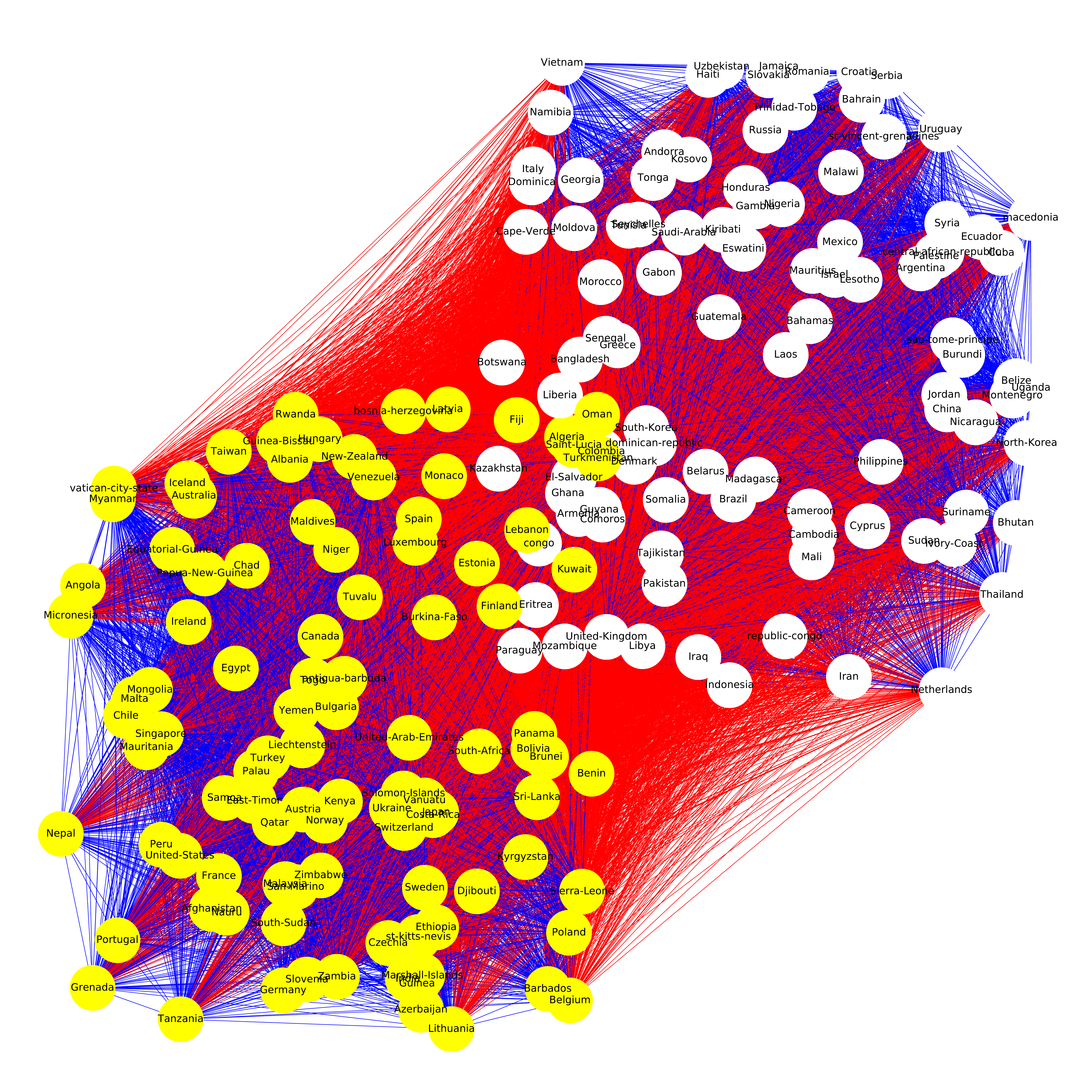}
\textbf{    \caption{Graphical visualization of Coalitions formed by proposed SSN model} }
\label{fig:my_label13}
    \end{center}
\end{figure*}
\subsubsection{Evaluation metrics}As explained in the section 6.4 of this paper, we needed a well known allies and enemies set of the real world, to evaluate which predicted coalition has the highest score. Table \ref{table:table6} shows the data collected for the same. 
The allies and enemies are captured through various sources including foreign affairs websites of different countries.
These sources are mentioned against each pair in Table VII.


\begin{figure*}[!hpt]
    \begin{center}
    \includegraphics[width=16cm,height=30cm,keepaspectratio]{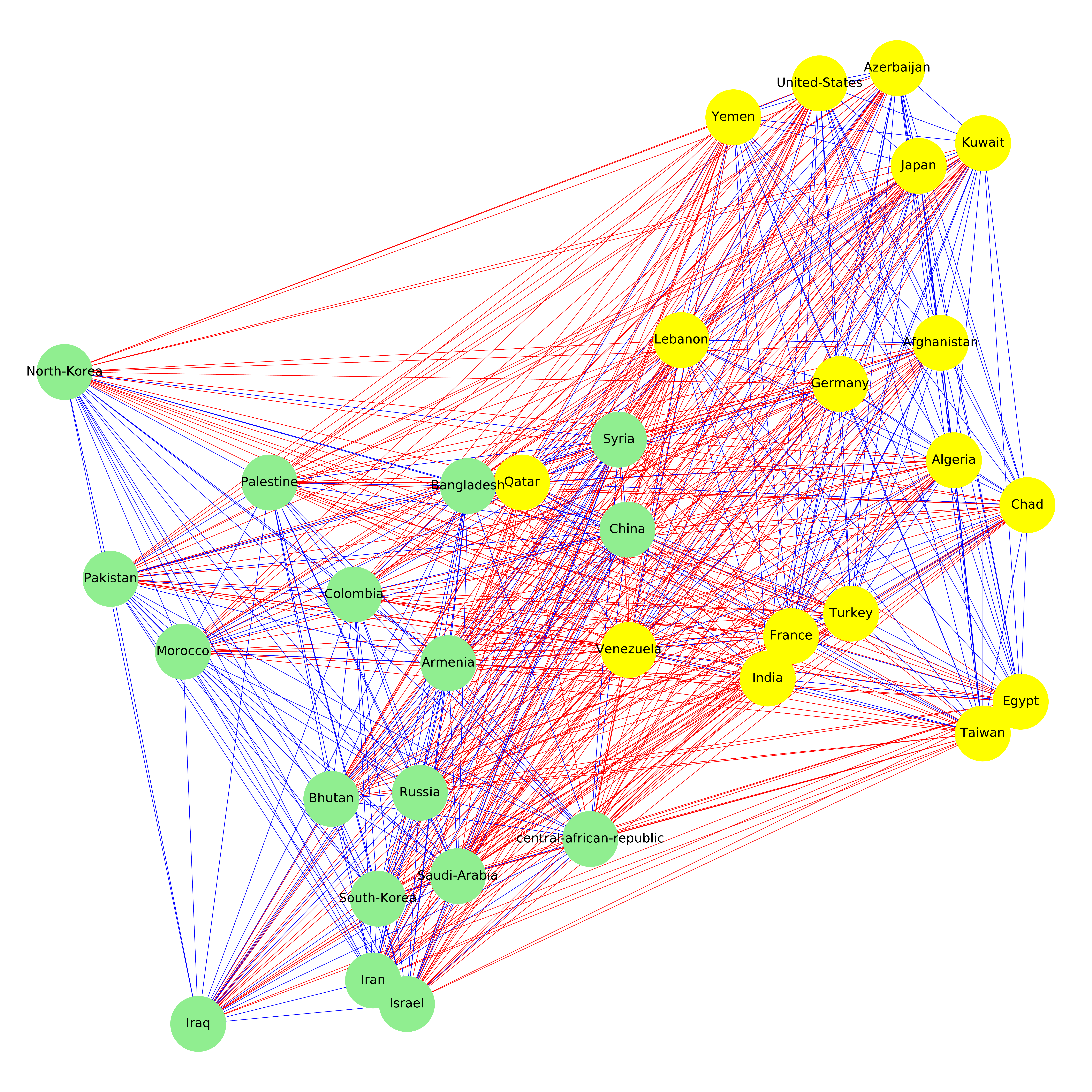}
\textbf{    \caption{Visual Coalitions graph for subset of countries (for better visualization)} }
    \label{fig:my_label14}
    \end{center}
\end{figure*}
\begin{minipage}{0.90\linewidth}
\begin{table}[H]
\small
\begin{center}
\begin{tabular}{||c  c ||}
\hline
Enemies & Allies \\ [0.5ex] 
\hline\hline
\hspace{0.5mm}
\begin{minipage}{0.45\linewidth}
\vspace{2mm} 
  (Pakistan, Afghanistan), (Pakistan, Bangladesh), (Pakistan, India), (Pakistan, United-States), (China, United-States), (China, India), (China, Taiwan), (China, Japan), (Russia, United-States),  (North-Korea, United-States), (North-Korea, South-Korea), (Turkey, Armenia), (Azerbaijan, Armenia), (Algeria, Morocco), (Israel, Lebanon), (Israel, Syria), (Israel, Egypt), (Israel, Palestine), (Israel, Yemen), (Iraq, Lebanon), (Iraq, Qatar), (Iraq, Kuwait), (Saudi-Arabia, Qatar), (Colombia, Venezuela), (central-african-republic, Chad)
\end{minipage}%
\hspace{0.5mm}
\vspace{1mm}
&
\begin{minipage}{0.40\linewidth}
\begin{tabular}{|p{\linewidth}}
  (Turkey, Azerbaijan), (India, United-States), (Japan, United-States),(Iraq, United-States), (Japan, Bangladesh), (Iran, Russia), (Iran, Iraq), (Iran, Venezuela), (India, Afghanistan), (Israel, United-States), (Israel, Iraq), (Germany, United-States), (Germany, France), (India, Russia), (India, Japan), (India, Bhutan), (Pakistan, Turkey), (Saudi-Arabia, China)
\end{tabular}
\end{minipage}%
\hspace{1.75mm}
\\
\hline
\end{tabular}
\textbf{\caption{\label{table:table6}Real-World known pairs of enemies and allies for evaluation. } }
\end{center}
\end{table}
\end{minipage}
\subsection{Interpretation of findings}
\vspace{2.9pt}
A detailed evaluation of the predicted sets versus The expected results from the real-world data as shown in Table \ref{table:table6} can be seen in Table \ref{table:table7}. From the total of 43 such pairs, The prediction sets discussed in Table \ref{table:table5} get a positive score of 32 with only 11 mismatches to the Ideal Arrangement. The Major mismatches are (North-Korea,South-Korea), (Israel,Palestine) and (Israel,United-States).
\begin{table}[hpt!]
\begin{minipage}{0.4\textwidth}
\begin{center}
\footnotesize
\begin{tabular}{||p{3.35cm} p{1.25cm} p{1.25cm} p{1cm}||}
\hline
\textbf{Nation-pair} & \textbf{Evaluation Data-set} & \textbf{Predicted Set} & \textbf{Right/ Wrong}\\ 
[0.35ex]
\hline\hline
Pakistan,Afghanistan \cite{enwiki:AfghanPak} & Opposite & Opposite & \Checkmark \\
\hline
Pakistan,Bangladesh \cite{enwiki:BangPak} & Opposite & Same & \XSolid \\
\hline
Pakistan,India \cite{enwiki:INdiaPak} & Opposite & Opposite & \Checkmark\\
\hline
Pakistan,United-States \cite{enwiki:PakUS}& Opposite & Opposite & \Checkmark\\
\hline
China,United-States \cite{enwiki:ChinaUS} & Opposite & Opposite & \Checkmark\\
\hline
China,India \cite{enwiki:ChinaIndia} & Opposite & Opposite & \Checkmark\\
\hline
China,Taiwan \cite{enwiki:ChinaTaiwan} & Opposite & Opposite & \Checkmark\\
\hline
China,Japan \cite{enwiki:ChinaJapan} & Opposite & Opposite & \Checkmark\\
\hline
Russia,United-States \cite{enwiki:RussiaUS} & Opposite & Opposite & \Checkmark\\
\hline
Iraq,United-States \cite{enwiki:IraqUS} & Same & Opposite & \XSolid\\
\hline
North-Korea,United-States \cite{enwiki:NorthKoeaUS} & Opposite & Opposite & \Checkmark\\
\hline
North-Korea,South-Korea \cite{enwiki:NorthKoreaSOuthKorea}& Opposite & Same  & \XSolid \\
\hline
Turkey,Armenia \cite{enwiki:ArmeniaTurkey}& Opposite & Opposite & \Checkmark\\
\hline
Azerbaijan,Armenia \cite{enwiki:ArmeniaAzerbejian}& Opposite & Opposite & \Checkmark\\
\hline
Algeria,Morocco \cite{enwiki:AlgeriaMorooco}& Opposite & Opposite & \Checkmark\\
\hline
Israel,Lebanon \cite{enwiki:IsraelLebanon}& Opposite & Opposite & \Checkmark\\
\hline
Israel,Syria \cite{enwiki:IsraelSyria} & Opposite & Same & \XSolid \\
\hline
Israel,Egypt \cite{EgyptIsrael}& Opposite & Opposite & \Checkmark\\
\hline
Israel,Palestine \cite{enwiki:IsraelPalestine} & Opposite & Same & \XSolid \\
\hline
Israel,Yemen \cite{enwiki:IsraelYemen} & Opposite & Opposite & \Checkmark\\
\hline
Israel,Iraq \cite{enwiki:IraqIsrael}& Same & Same & \Checkmark \\
\hline
Iraq,Lebanon \cite{enwiki:IraqLebanon} & Opposite & Opposite & \Checkmark\\
\hline
Iraq,Qatar \cite{enwiki:IraqQatar} & Opposite & Opposite & \Checkmark\\
\hline
Iraq,Kuwait \cite{enwiki:Iraqkuwait} & Opposite & Opposite & \Checkmark\\
\hline
Saudi-Arabia,Qatar \cite{enwiki:QatarSaudiaArabia} & Opposite & Opposite & \Checkmark\\
\hline
Colombia,Venezuela \cite{ColombiaVenzula}& Opposite & Opposite & \Checkmark\\
\hline
central-african-republic,Chad \cite{ChadAnfCAfricaRepublic}& Opposite & Opposite & \Checkmark\\
\hline
Turkey,Azerbaijan \cite{enwiki:AzerbeajistanTurkey}& Same & Same & \Checkmark\\
\hline
India,United-States \cite{enwiki:INDiaUS}& Same & Same & \Checkmark\\
\hline
Japan,United-States \cite{enwiki:JapanUS} & Same & Same & \Checkmark\\
\hline
Japan,Bangladesh \cite{enwiki:japanBangla} & Same & Opposite & \XSolid \\
\hline
Iran,Russia \cite{enwiki:IranRussia}& Same & Same & \Checkmark\\
\hline
Iran,Iraq \cite{IranIraq}& Same & Same & \Checkmark\\
\hline
Iran,Venezuela \cite{enwiki:IranVenzula}& Same & Opposite & \XSolid \\
\hline
India,Afghanistan \cite{enwiki:AfghanIndia} & Same & Same & \Checkmark\\
\hline
Israel,United-States \cite{enwiki:IsraelUS} & Same & Opposite & \XSolid \\
\hline
Germany,United-States \cite{enwiki:GermanyUS} & Same & Same & \Checkmark\\
\hline
Germany,France \cite{enwiki:FranceGermany} & Same & Same & \Checkmark\\
\hline
India,Russia \cite{enwiki:IndiaRUssia}& Same & Opposite & \XSolid \\
\hline
India,Japan \cite{enwiki:IndiaJapan} & Same & Same & \Checkmark\\
\hline
India,Bhutan \cite{enwiki:BhutanINdia} & Same & Opposite & \XSolid \\
\hline
Pakistan,Turkey \cite{enwiki:PakistanTurkey} & Same & Opposite & \XSolid \\
\hline
Saudi-Arabia,China \cite{ChinaSaudiaArabia} & Same & Same & \Checkmark\\
\hline
\end{tabular}
\textbf{\caption{\label{table:table7}Evaluation data-set vs prediction set values} }
\end{center}
\end{minipage}
\end{table}
From our experiment on cross validation against manual scavenged pair of allies and enemies our proposed model gave 32 correct pairs against 43 total pairs. Thus we have gained an accuracy of 74 to 75 percent in achieving our goal.
\section{Conclusion}
In this era where every single country wants to gain power as well as be independent in its own successful terms, the thrust to defeat others and stand as superpower prevails. The notion of war rises  due to hatred, insecurity, curiosity, economic gain, territorial gain, religion, nationalism, revenge etc. It can be analyzed that pertaining to the current temperament of the parliaments of each country, World War-III might happen. Even though it is not in our hands to halt it, utmost we can do is try to predict what can be the possibilities in wars. Through our paper we have chosen one of the most inquisitive topics of interest that is “The World War-III-Coalitions” where we have answered and predicted two of the cunning happening of WW-III. In a graph representation of participating nodes in WW-III, we have predicted the best possible coalitions of countries that will be formed. We have also extended analysis on various parameters influencing the prediction strategies through graph representation and have created evaluation matrices to compare our predicted results with the real world results. Also, we have prepared a data-set that consist of country pairs and various attributes that pertains between them that influenced our prediction strategies. 
The future direction for the work can be use of link prediction in social networks and use of other coalitions finding algorithms. Other open directions can be use of more non linear functions for score calculation and try different hyper-tuning techniques like grid search or neural networks.
\section*{Acknowledgement}
This work has been done as a part of the coursework at IIT Ropar. We would like to thank our instructor, 'Dr. Sudarshan Iyengar', for his unique teaching methodology and motivation behind this paper.
\bibliographystyle{IEEEtran}
\bibliography{bib1}

\end{document}